%% file: SwiftJ1357_XMM.tex
\documentclass[useAMS,usenatbib]{mn2e}

\usepackage{epsfig}
\usepackage{amsmath}
\usepackage{amssymb}
\usepackage{natbib}
\usepackage{threeparttable} 

\usepackage{rotating}

\usepackage[hyphens]{url}
\usepackage{hyperref}

\usepackage{epstopdf}

\usepackage{booktabs,array}

\usepackage{color}

\newcommand{\swiftj}{Swift J1357.2--0933}

\include{def_bibtex}

\include{definitions}

\title[An X-ray view of \swiftj]{An X-ray view of the very faint black hole X-ray transient \swiftj\ during its 2011 outburst }
\author[M. Armas Padilla et al.]
{M. Armas Padilla$^{1}$\thanks{e-mail: m.armaspadilla@uva.nl}, R. Wijnands$^{1}$, D. Altamirano$^{1}$, M. M\'endez$^{2}$, J. M. Miller$^{3}$ \newauthor and N. Degenaar$^{3}$\thanks{Hubble Fellow} \\
$^{1}$Astronomical Institute ``Anton Pannekoek", 
University of Amsterdam, 
Postbus 94249, 1090 GE Amsterdam, The Netherlands\\
$^{2}$Kapteyn Astronomical Institute, University of Groningen, Postbus 800, 9700 AV Groningen, The Netherlands\\
$^{3}$Department of Astronomy, University of Michigan, 500 Church Street, Ann Arbor, MI 48109, USA\\
}

\begin{document}

\date{DRAFT VERSION}

\pagerange{\pageref{firstpage}--\pageref{lastpage}} \pubyear{0000}

\maketitle

\label{firstpage}

\begin{abstract} 
We report on the X-ray spectral (using \xmm\, data) and timing
behavior (using \xmm\, and {\it Rossi X-ray Timing Explorer} [\rxte]
data) of the very faint X-ray transient and black hole system
\swiftj\, during its 2011 outburst. The \xmm\ X-ray spectrum of this
source can be adequately fitted with a soft thermal component with a
temperature of $\sim0.22$~keV (using a disc model) and a hard,
non-thermal component with a photon index of $\Gamma\sim$1.6 when
using a simple power-law model. In addition, an edge at $\sim0.73$~keV
is needed likely due to interstellar absorption. During the first
\rxte\, observation we find a 6 mHz quasi-periodic oscillation (QPO)
which is not present during any of the later \rxte\, observations or
during the \xmm\, observation which was taken 3 days after the first
\rxte\ observation. The nature of this QPO is not clear but it could
be related to a similar QPO seen in the black hole system H 1743--322
and to the so-called 1 Hz QPO seen in the dipping neutron-star X-ray
binaries (although this latter identification is quite
speculative). The observed QPO has similar frequencies as the optical
dips seen previously in this source during its 2011 outburst but we
cannot conclusively determine that they are due to the same underlying
physical mechanism. Besides the QPO, we detect strong band-limited
noise in the power-density spectra of the source (as calculated from
both the \rxte\, and the \xmm\, data) with characteristic frequencies
and strengths very similar to other black hole X-ray transients when
they are at low X-ray luminosities. We discuss the spectral and timing
properties of the source in the context of the proposed very high
inclination of this source. We conclude that all the phenomena seen
from the source cannot, as yet, be straightforwardly explained neither
by an edge-on configuration nor by any other inclination configuration
of the orbit.

\end{abstract}

\begin{keywords}
accretion, accretion discs - 
stars: individuals (\swiftj) - 
stars: black hole - 
X-rays: binaries
\end{keywords}


\section{Introduction}\label{sec:intro}

In our Galaxy, the brightest X-ray emission originating from point
sources is produced by binary systems harbouring a black hole or
neutron star that is accreting material from a (sub)solar mass
companion star. However, the X-ray luminosity exhibited by these low
mass X-ray binaries (LMXBs) can vary by several orders of magnitude,
in particular in the so-called X-ray transients. Those sources are a
subgroup of LMXBs that spend most of their time in a quiescent state
with a low X-ray luminosity ($\lx\sim10^{30}-10^{33}\lum$), and
sporadically transit to an outburst episode, in which the accretion
rate increases drastically, resulting in several orders of magnitude
increase in their X-ray brightness. These systems can be classified
based on their 2--10~keV peak luminosity ($L_X^{peak}$) reached during
outbursts. Very faint X-ray transients \citep[VFXTs;][]{Wijnands2006}
are those systems that reach $\lx^{peak}$ of only
$\sim10^{34}-10^{36}\lum$, orders of magnitude lower than the
brighter, regular systems (which display
$\lx^{peak}\sim10^{36}-10^{39}\lum$).

The low X-ray luminosities in combination with the duration of their
outburst and quiescent episodes (i.e., duty cycles) imply that the
VFXTs accrete at very low rates, both instantaneously as well as
averaged over many outbursts-quiescence cycles
\cite[][]{Degenaar2009,Degenaar2010b}. In fact, some of these
sub-luminous sources might have such low long-term time-averaged
inferred mass-accretion rates that exotic evolution scenarios are
necessary to explain their existence
\citep[][]{2006MNRAS.366L..31K,Wijnands2006,Degenaar2009}. A large
fraction of the VFXTs has conclusively been identified as accreting
neutron star systems by the detection of thermonuclear X-ray bursts
\citep[e.g., ][]{Maeda1996, Cornelisse2002, Chelovekov2007,
DelSanto2007, Wijnands2009,Degenaar2010b} but for the majority of
VFXTs, the nature of the accretor remains elusive.
  
\begin{figure}
\begin{center}
\includegraphics[angle=0,width=\columnwidth]{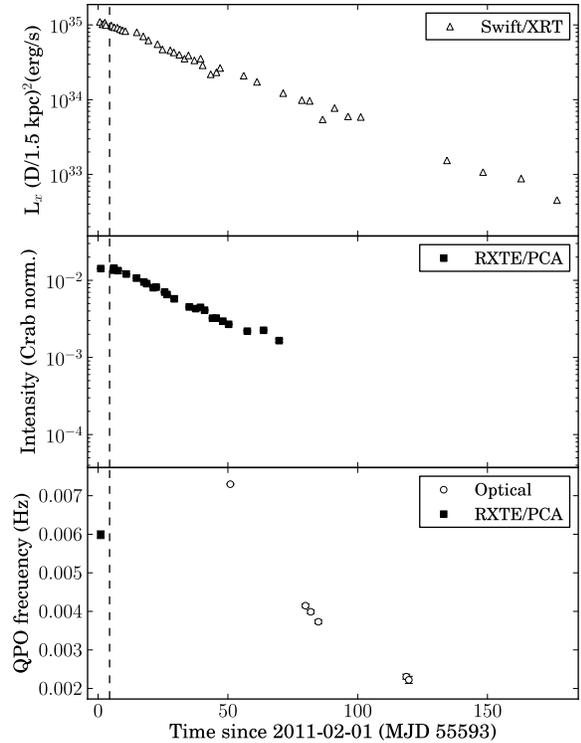}\\
\caption{The \swift/XTE light--curve ({\it top panel}; from \citealt{ArmasPadilla2013}) and the {\it RXTE}/PCA light--curve ({\it middle panel}) of the outburst of Swift J1357.2--0933. {\it Bottom panel:} The evolution of the frequency of the mHz QPOs seen in {\it RXTE} data (black square) and in the optical (white circles; from \citealt{Corral-Santana2013}). The dashed--line indicates when the {\it XMM-Newton} observation was performed. }
\label{fig:lc_qpo}
\end{center}
\end{figure}

\swiftj\ is so far the only confirmed black hole VFXT
\citep[][]{ArmasPadilla2013,Corral-Santana2013}. The source was
discovered in outburst on January 28, 2011 with the Swift Burst Alert
Telescope (BAT; \citealt{Barthelmy2005, Krimm2011}). Subsequent
observations were carried out with several X-ray satellites as well as
ground-based telescopes \citep[e.g.,
][]{Krimm2011b,Milisavljevic2011,Torres2011, Casares2011}. The X-ray
outburst lasted $\sim7$ months (e.g., see Fig.~\ref{fig:lc_qpo}, top
panel) and reached a 2--10~keV X-ray peak luminosity of
$L_X^{peak}\sim10^{35}~\lum$ which classified the system as VFXT
\citep{ArmasPadilla2013}.  Photometry of the possible counterpart
found in archival Sloan Digital Sky Survey (SDSS) data (at times when
the source is assumed not to be active) indicated that the quiescent
optical counterpart is very red and that the companion is likely an M4
star. From this, the distance toward the source was estimated to be
$\sim$1.5 kpc \citep[][]{Rau2011,Corral-Santana2013}.

Its short distance and the low Galactic extinction toward the source
(due to its location above the Galactic plane; b=50.0042), made it
possible to obtain good UV/optical observations during outburst in
addition to the X-ray observations. From optical data obtained during
outburst, \citet[][]{Corral-Santana2013} established the black hole
nature for the accretor (they obtained a minimum mass of the accretor
of $M_{x}>3.6\sol$) and they found a 2.8~hours orbital period. In the
same work they presented the detection of intense dips in the optical
light curve, which were explained by a toroidal structure in the inner
region of the disc seen at high inclination ($i\ga70\degr$) which is
moving outward as the outburst progresses.

Optical and infrared observations obtained in 2012 and 2013 when the
source was in quiescence \citep[][]{Shahbaz2013}, revealed similar
dips as seen during outburst as well as flare events. The light-curves
were dominated by a larger variability than what is observed in other
systems at similar time-resolution, which suggests that another source
of light with large variability completely dominates the optical flux
of the companion star (the companion star could not be
detected). \citet[][]{Shahbaz2013} estimated that the magnitude of the
companion start is $V_{min}=22-25$, which, combined with the expected
magnitude for a M4.5 V secondary star, resulted in a inferred distance
between 0.5~kpc and 6.3~kpc.

Using the X-ray and UV/optical data obtained with \swift\ during the
2011 outburst of \swiftj, \citet{ArmasPadilla2013} found that the
X-ray spectrum of the source softened with decreasing source
luminosity and that the X-ray flux was correlated with the UV and
optical magnitudes in such a way that it is likely that the black hole
is accreting via a non irradiated or only marginally irradiated disc.

In this paper, we present the spectral analysis of our \xmm\
observation of \swiftj\ obtained a few days after the peak of the
X-ray outburst. We also present a timing study using the same \xmm\
data in combination with data obtained throughout the outburst using
the {\it Rossi X-ray Timing Explorer} (\rxte).

\section{Observations and analysis}\label{sec:Obs}


\subsection{\rxte\ observations}\label{subsec:rxte}

We used data from the \rxte\ Proportional Counter Array \citep[PCA;
for instrument information see][]{1993SPIE.2006..324Z}. There were 26
pointed observations (all ObsID starting with 96065-02 and 96419-01),
each consisting of a fraction of one to three entire satellite
orbits. We used the 16-s time-resolution Standard 2 mode data to
calculate the 2--20 keV intensity as described in
\citet{2008ApJ...685..436A}.  We subtracted the background and
corrected for the deadtime. In order to correct for differences in
effective area between the PCUs themselves, we normalized our
intensity by the corresponding Crab color values \citep[see][for more details about this method]{2008ApJ...685..436A}. We used all active
PCUs to calculate the average intensity per observation. The results
are plotted in Fig.~\ref{fig:lc_qpo} (middle panel).

For the Fourier timing analysis we used the data obtained using the
Good Xenon mode, which has a time resolution of $\sim
1\mu$s. Leahy-normalized power density spectra (PDS) were constructed
using data segments of both 128 and 512 s.  No background or
deadtime corrections were made prior to the calculation of the power
spectra.  We first averaged the power spectra per observation. We
inspected the shape of the average power spectra at high frequency (
$>2000$~Hz) for unusual features in addition to the usual Poisson
noise but none were found.  We then subtracted the Poisson noise by
estimating it from the power between 3000 and 4000~Hz, where neither
intrinsic noise nor QPOs are known to be present, using the method
developed by \citet{2004PhDT.......415K} based on the analytical
function of \citet{1995ApJ...449..930Z}.  The resulting power spectra
were converted to squared fractional rms \citep{1990A&A...227L..33B,1991ApJ...383..784M,1995lns..conf..301V}.
In this normalization the square root of the integrated power density
equals the variance of the intrinsic variability in the source count
rate.

\subsection{\xmm\ observation}\label{subsec:xmm}

The \xmm\ observatory \citep{Jansen2001} pointed at \swiftj\ for
$\sim37$~ks on 2011 February 5 (Obs ID 0674580101). In the current
paper, we are using the data obtained with the European Photon Imaging
Camera (EPIC) which contains two MOS cameras \citep{Turner2001} and
one PN camera \citep{Struder2001}, as well as the data from the
Reflection Grating Spectrometers \citep[RGS,][]{Hender2001}. During
the observation the medium optical blocking filter was used. The MOS1
camera was operated in the small window imaging mode, while the MOS2
and PN cameras were operated in timing mode. To reduce the data we
used the Science Analysis Software (SAS, v. 12.0).

In order to omit episodes of background flaring we excluded the data
where the high-energy count rate was larger than 0.25~$\cnts$
($>$10~keV) and 0.6~$\cnts$ (10-12~keV) for the MOS and PN cameras,
respectively, which resulted in a total live time of $\sim29$~ks. We
used the task EPATPLOT to assess the presence of pile--up in our
data. MOS1 was strongly affected (the source count rate is
$\sim25~\cnts$ when the pile-up starts to be an issue at count rates
$>5~\cnts$). In order to avoid the pile-up it is necessary to
eliminate a large central portion (annular extraction region with a
13\arcsec inner radius and a 40\arcsec outer radius) which results in
a very low statistic spectrum. Therefore we decide to exclude the MOS1
data from our analysis. We also excluded the MOS2 data because of the
large uncertainties which still exist in the calibration of timing
data obtained using this instrument.

We extracted the PN source and background events using the RAWX
columns in [30:45] and in [2:9], respectively. The net source count
rate is $\sim150~\cnts$ (0.4-11~keV), which is much lower than the
$\sim800~\cnts$ at which the pile--up starts to affect the
observation. However, the EPATPLOT tool suggests that the observation
is affected by pile--up and it is necessary to exclude the three
central columns (RAWX columns in [37:39]) to mitigate the effects. We
generated the spectrum for both column selections and compared them,
and found that the two spectra are fully consistent with each other. Therefore, we
concluded our observation was not affected or at most only slightly
affected by pile--up and do not exclude any central columns from our
source selection.  We have also compared the spectrum when the
background is subtracted and when it is not subtracted. When the PN
CCD is operated in timing mode, bright sources might illuminate the
entire CCD so that background spectra will be contaminated by photons
from the source \citep[e.g.,][]{Dunn2010,Hiemstra2011}. However, there
is no spectral shape difference between the background-subtracted
spectrum and the one without background subtraction. For completeness,
we still have corrected for the background in our analysis. We
generated the light curves and the spectra, as well as the associated
response matrix files (RMFs) and ancillaries response files (ARFs)
using the standard analysis
threads\footnote{\url{http://xmm.esac.esa.int/sas/current/documentation/threads/}}.

We used the \textsc{SAS} task \textsc{rgsproc} to reduce the RGS data
and to produce the response matrices and spectra. We selected the
option keepcool=no and withrectification=yes. The first option
excludes columns that give signals a few percent below the values
expected from their immediately neighbours and are only likely to be
relevant when studying weak absorption features in spectra with high
statistics \citep[see \textsc{SAS} User guide section
5.6.3;][]{DiazTrigo2012}. The second option activates the current
calibration file (CCF) that causes the empirical rectification
correction factors, to be applied to RGS effective area to bring
simultaneous models of RGS and EPIC-PN spectra into agreement by
reducing apparent systematic errors to the level of better than about
1\% (see the XMM-Newton CCF release note XMM-CCF-REL-269).

\subsubsection{Spectral analysis}

All spectra were grouped to contain a minimum of 25 photons per bin
and the energy resolution was not oversampled by more than a factor
3. We fitted all spectra using \textsc{XSpec}
\citep[v. 12.8;][]{xspec}. We included the photoelectric absorption
component (\textsc{phabs}) to account for the interstellar absorption
assuming the cross--sections of \citet{verner1996} and the abundances
of \citet{Wilms2000}.

We combined the 0.4--1.8~keV RGS spectra (RGS1 and RGS2 first order)
and the 0.7--10~keV EPIC-PN spectrum and fitted them simultaneously
with the parameters tied between the three detectors. We added a
constant factor (\textsc{constant}) to the spectral models with a
value fixed to 1 for EPIC-PN spectrum and free to vary for the RGS
spectra in order to account for cross calibrations uncertainties. We
added a 1\% systematic errors to all instruments to account for the
errors in the relative calibration between the RGS and EPIC-PN
detectors \citep[][XMM-SOC-CAL-TN-0052]{Kirsch2004}.


\begin{table}
\caption{Best fit to the 0.4--1.8~keV RGSs spectra and 0.7--10~keV PN spectrum using \textsc{phabs*(diskbb+nthcomp)*edge}.}
\label{tab:results}
\begin{threeparttable}
\begin{tabular}{l c }
\hline \hline
Parameter [unit] &  \\
\hline
\multicolumn{2}{c}{\textsc{phabs}}\\
\Nh[$\times10^{12}\nh$] & 3.52 \\ 
\multicolumn{2}{c}{\textsc{diskbb}}\\
$kT_{\rm in}$[keV] & $0.216 \pm 0.005 $\\ 
$N_{\rm Diskbb}$ & $1176 \pm 70$\\ 
\multicolumn{2}{c}{\textsc{nthcomp}}\\
$\Gamma$ & $1.61 \pm 0.01 $ \\
$kT_{\rm e}$ [keV] & $8.23^{+1.57}_{-1.04}$\\
$kT_{\rm bb}$ [keV] &  0.216 (tied)   \\
$N_{\rm comp}$ &  $0.0420\pm 0.0003$  \\
\multicolumn{2}{c}{\textsc{edge}}\\
edgeE [keV] & $0.731 \pm 0.005$\\
MaxTau    &  $0.10 \pm 0.01$  \\
\hline
\Fx,$_{\rm abs}$ [$\times10^{-10}~\flux$] & $3.34 \pm 0.04  $\\
\Fx,$_{\rm unabs}$ [$\times10^{-10}~\flux$] & $3.34\pm 0.04 $\\
$\lx$ [$\times10^{34}~\lum$] &  $8.99 \pm 0.04$\\
Thermal fraction & 6.6\% \\
\hline
\chis (dof) & 1.08 (1136) \\
\hline
\end{tabular}
\begin{tablenotes}
\item[Note]{To calculate the parameter errors, we fixed the instrumental features components. }
\end{tablenotes}
\end{threeparttable}
\end{table}

\subsubsection{Timing analysis}

Leahy-normalized power spectra were constructed (in segments of
$\sim$537 s; no background or deadtime corrections were applied
during this) using the \xmm/PN data in the 0.5--10 keV energy range
and using the same source extraction region as we used during the
spectral analysis. We added all power spectra together to get an
averaged one for the whole observation. We subtracted the Poission
noise spectrum, which was estimated for frequencies between 100 and
200 Hz. The power spectrum was converted to squared fractional rms in
the same way as the {\it RXTE} data (see Section~\ref{subsec:rxte}).

\section{Results}

\subsection{Spectral results}


\begin{figure}
\begin{center}
\includegraphics[angle=-90,width=\columnwidth]{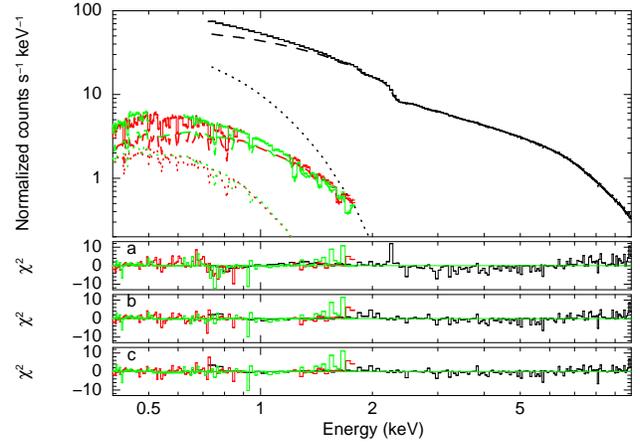}\\
\caption{ PN (black), RGS1 (red) and RGS2 (green) spectra. The solid lines represent the best fit for a combined \textsc{nthcomp} (dashes), \textsc{diskbb} (dotted lines) and \textsc{edge} (at $\sim0.73$~keV) model. In the sub-panels the contributions to \chis\ are plotted  when fitting with a) \textsc{phabs*(diskbb+po)} (without instrumental features corrections), b) \textsc{phabs*(diskbb+po)*edge} and c)  \textsc{phabs*(diskbb+nthcomp)*edge}.}
\label{fig:spec}
\end{center}
\end{figure}


The first model which we used to fit the \xmm\ spectrum was a
power--law (\textsc{powerlaw}) model with an accretion disc model
consisting of multiple blackbody components
\citep[\textsc{diskbb};][]{Makishima1986}. With a \chired\ of 1.3 for
1146 degrees of freedom (dof) the fit was not acceptable (with a
p-value\footnote{The p-value represents the probability that the
deviations between the data and the model are due to chance alone. In
general, a model can be rejected when the p-value is smaller than
0.05.} of $\sim 1.9 \times 10^{-9}$). The residuals
(Fig.~\ref{fig:spec}(a)) showed an emission feature at $\sim2.2$~keV,
likely produced by a over-compensation by the CTI correction in the
characteristic EPIC-PN residual due to the Gold (2.21~keV) edge
(Kolehmainen et al. 2013, in prep.). We added a Gaussian component
(\textsc{gauss}) in our models to mitigate this instrumental
feature. Additionally, the RGS spectra show residuals at
$\sim0.54$~keV and $\sim0.65$~keV energies, possibly due to
instrumental inefficiencies at the oxygen and iron edges, respectively
(see the RGS calibration status XMM-SOC-CAL-TN-0030), and which were
not accurately corrected in the calibration process. Therefore, we
included two components (\textsc{edge}) in our fits to model such
features. From now on, unless otherwise mentioned, we include these
instrumental feature components in all our models.

The residuals also show an absorption feature at $\sim0.73$~keV. The
interpretation for this feature can be absorption by interstellar iron
L2 ($\sim0.73$ keV) or L3 ($\sim0.71$ keV)
\citep[][]{Wilms2000,DiazTrigo2007, Pinto2013}. We incorporated an
\textsc{edge} component to reproduce this feature in our models.

Once we had added these additional components to model the residuals
(Fig.~\ref{fig:spec}(b)), the fit with the \textsc{diskbb+powerlaw}
model improves considerably (\chired$\sim 1.13$ for 1137 dof) although
still it is not an acceptable fit (with a p-value of $\sim$0.002). The
disc temperature obtained was $\sim0.25$~keV and the photon index of
the power-law component was $\Gamma\sim1.51$. The photon index is in
agreement with the values reported by \citet{ArmasPadilla2013} in
their analysis of the quasi--simultaneous \swift\ data of the source
($\sim$1~day of difference between the \xmm\ and \swift\ observations). The unabsorbed
0.5--10~keV flux is $3.3\times10^{-10}\flux$, to which the thermal
component contributes $\sim$7\%. Assuming a distance of 1.5~kpc the
inferred X-ray luminosity is $\lx\sim8.9\times10^{34}\lum$.

We repeated the fit replacing the hard \textsc{powerlaw} component
with a thermally comptonized continuum model
\citep[\textsc{Nthcomp};][]{Zdziarski1996, .Zycki1999}. We tied the
seed photon temperature (low energy rollover; $kT_{\rm bb}$) to the diskbb
temperature ($T_{\rm in}$). This change in the model to fit the hard
component slightly reduces the \chired\ to 1.08 for 1136 dof (with a
p-value of $\sim$0.03). The power-law photon index is $\Gamma\sim1.6$,
the diskbb temperature $\sim0.22$~keV and the electron temperature
(high energy rollover; $kT_{\rm e}$) is $\sim8$~keV. The flux obtained
using this model is the same than with the power--law component.  We
also repeated the fits replacing the \textsc{diskbb} with the
\textsc{diskpn} \citep{Gierlinski1999}, which is an extension of the
\textsc{diskbb} model that includes corrections for temperature distribution
near the black hole. However, the results were nearly identical when
using the \textsc{diskbb} model.

The impossibility to decrease the \chired\ and get an acceptable fit
could be due to remaining uncertainties in the cross-calibration
between the different detectors (e.g. we can see in
Fig.~\ref{fig:spec} that there is a big contribution to the \chis\ in
the overlapping between the EPIC-PN and RGS2 cameras between
1--2~keV). The results of the best spectral analysis
(\textsc{phabs*(diskbb+nthcomp)*edge}) are reported in
Table~\ref{tab:results}. The uncertainties on the spectral parameters
are at 90\% confidence level and the flux errors have been calculated
following the procedure presented by \citet{Wijnands2004}.

\subsection{Timing results}

\begin{figure}
\includegraphics[angle=0,width=8cm]{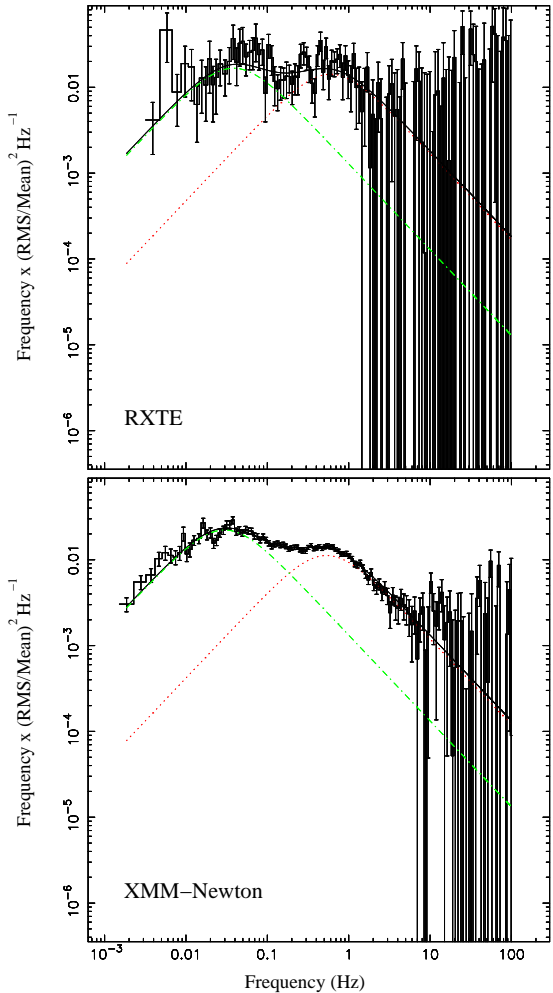}\\
\caption{{\it Top panel:} Broad-band PDS (2--20 keV) of the first {\it RXTE} observation (ObsID: 96065-02-01-00; 2 Feb 2011). Clearly a strong broad-band noise is present, but below 0.01 Hz also the mHz QPO is visible. {\it Bottom panel:} Broad-band PDS (0.5--10 keV) of the {\it XMM-Newton}/PN observation performed on 5 Feb 2011. This also clearly shows the strong broad-band noise, but no evidence for the mHz QPO. In both panels, the dotted and dashed-dotted lines are the best fit model assuming two Lorentzians. \label{fig:pds}}
\end{figure}

\begin{figure}
\includegraphics[angle=0,width=6cm,angle=-90]{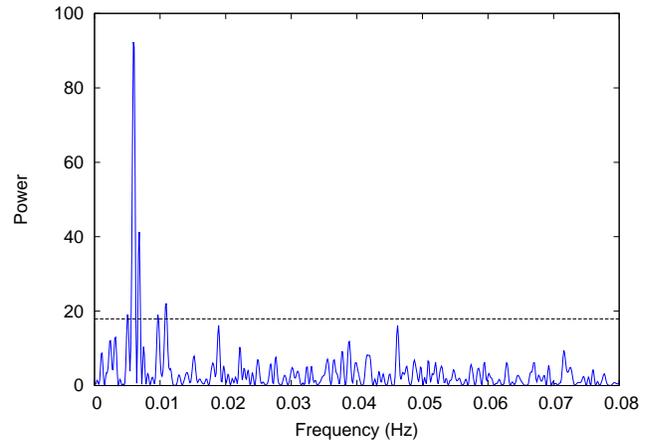}\\
\caption{The Lomb-Scargle diagram of the {\it RXTE} observation performed on 2 Feb 2011 (ObsID: 96065-02-01-00) clearly showing the mHz QPO. The dashed line is the 4 $\sigma$ significance level (assuming no red noise contribution). \label{fig:qpo}}
\end{figure}

The power spectrum of the first {\it RXTE} observation (ObsID
96065-02-01-00; $\sim$~2~ks obtained on 2 Feb 2011) is shown in Fig.~\ref{fig:pds}
(top panel). Clearly a strong broad-band noise component is present
which has a total rms amplitude (between 0.01 and 100 Hz) of
35$\pm$4 \% (2--20 keV). We fitted the PDS using two zero-centered Lorentzians \citep{Belloni2002}. 
The obtained characteristic frequencies
and fractional rms amplitudes (2--20 keV) of the two components are
0.6$\pm$0.1 Hz and 0.039$\pm$0.007 Hz, and 21$\pm$1 \% and
23$\pm$1 \%, respectively. The power spectra obtained for the other
{\it RXTE} observations look very similar to this power spectrum, but
with decreasing quality because of the decreasing count rate as the
outburst progressed.

From Fig.~\ref{fig:pds}, top panel, it can be seen that in the PDS of
the first {\it RXTE} observation excess power is present at a
frequency of $\sim$0.005 Hz. The PDS shows indications of a peak at
$\sim$6 mHz. To investigate this feature further we used 1-s
resolution GoodXenon mode PCA light curves in the $\approx2-60$~keV
range and searched for periodicities in each {\it RXTE} orbit
separately using Lomb-Scargle periodograms
\citep{1976Ap&SS..39..447L,1982ApJ...263..835S}. The
Lomb-Scargle periodogram of the first {\it RXTE} observation (which was one orbit long) is shown
in Fig.~\ref{fig:qpo}, showing clearly a quasi-periodic oscillation
(QPO) at 5.9$\pm$0.1 mHz, with a FWHM $<$ 2 mHz (and thus quality
factor of $\sim$3 or larger). The rms amplitude (2--60 keV) of this
QPO is 12$\pm$3 \%. We also searched the other {\it RXTE}
observations for a similar frequency QPO, but none were found. However, given the low count rate and the length of the observations the upper limits are not constraining (the 3 sigma upper limits are typically of the order of tens of percent fractional rms). Since
the second {\it RXTE} observation (ObsID: 96065-02-02-00) was
performed on 7 Feb 2011, this means that the QPO lasted less than 5
days after the first \rxte\ observation.

The power spectrum of our \xmm\ data is shown in
Fig.~\ref{fig:pds} (bottom panel). Clearly a similar broad-band noise
(with a total 0.5--10 keV rms amplitude between 0.01 and 100 Hz of
33$\pm$2\%) is seen as in the {\it RXTE} data (Fig.~\ref{fig:pds},
top panel), but without the mHz QPO. The 3 sigma upper limit on the strength of a possible QPO is 7\% rms in the 0.5--10~keV energy range. Since the {\it XMM-Newton}
observation was performed $\sim$3.5 days after the {\it RXTE}
observation during which the mHz QPO was observed, it means that this
QPO lasted less than 3.5 days after the \rxte\ observation. We fitted
the {\it XMM-Newton} PDS also with two zero-centered Lorentzians (similar to what we had done for the {\it RXTE} data), which
provided an acceptable fit (\chired$ \sim 1.1$). The frequency and rms
amplitude of the two Lortenzians are 0.53$\pm$0.02 Hz and
0.030$\pm$0.001 Hz, and 18.8$\pm$0.2\% and 26.4$\pm$0.3\%,
respectively. We also fit the PDS with a broken-power law plus one Lorentzian model resulting in a break frequency of 0.038$\pm$0.002 Hz and a frequency of 0.33$\pm$0.06 Hz for the Lorentzian. Both fit models gave values which are very similar to what we have obtained
from the {\it RXTE} power spectra. We note that the {\it XMM-Newton}
data covers a different energy range (0.5--10 keV) than the {\it RXTE}
data (2--20 keV), therefore a direct comparison between the results
has to be done with care. A detailed timing study (e.g., evolution of
the power spectral features, energy dependency) is beyond the scope of
this paper.

\section{Discussion}

We report on the spectral and timing behaviour of the VFXT and black hole system \swiftj\, as observed with \xmm\ and on the timing  behaviour of the source as observed with \rxte. 

\subsection{Spectral behaviour}

Our high quality \xmm\ spectrum of \swiftj\ was acquired one week
after the source was discovered. The source was detected at a
luminosity of $\lx\sim9\times10^{34}(D/1.5~kpc)^{2}\lum$. The
best-fitting model to the spectrum consists of a thermal component
(\textsc{diskbb}) with a temperature of $kT\sim0.22$~keV, a hard
component (\textsc{nthcomp}) with a photon index of $\Gamma\sim1.6$
and kT$_{e}\sim$8.2~keV, and one \textsc{edge} located at
$\sim$0.73~keV.

The most probable origin for the thermal component is soft emission
from the accretion disc, given that the compact object is a black hole
\citep[][]{Corral-Santana2013}. The disc temperature obtained is low
compared to the typical 1-2~keV obtained for black hole X-ray
transients in their soft states
\citep[e.g.,][]{McClintock2006}. However, similar relatively cool
discs are observed in black hole systems in their low-hard state
\citep[e.g.,][]{Miller2006a, Miller2006b,Reis2010}.

The only clear spectral feature detected in the \swiftj\ spectrum is
an edge at $\sim$0.73~keV ($\sim17$\AA). This feature is likely due to
absorption by interstellar Fe (the L3 edge at $\sim$0.70~keV and L2
edge at $\sim$0.72~keV have been detected in other LMXBs; e.g.,
\citealt{DiazTrigo2007, Pinto2013}). But if indeed \swiftj\ is an
accretion disc corona (ADC) system seen in an edge-on geometry, as
suggested by \citet{Corral-Santana2013}, we would expect to see more
lines in our spectra intrinsic to the source (and not only due to
interstellar absorption). Searching in the literature for spectral
results of ADC sources, we found that generally they show a variety of
absorption and/or emission lines
\citep[e.g.,][]{Cottam2001,Kallman2003,2005A&A...436..195B,Iaria2013}.

Additionally, if \swiftj\ were an ADC source,
the intrinsic X-ray luminosity of the source would be approximately
one to two orders of magnitude higher than the luminosities we observe and therefore the peak of luminosity reached by the source would be
$\lx\sim10^{36-37}\lum$. However, \citet{ArmasPadilla2013}
found that the spectra of \swiftj\ became softer when the source
luminosity decreased. Such softening is observed in BH
systems at luminosities $\lx\lesssim10^{36}\lum$ \citep[e.g.,
][]{Plotkin2013,Corbel2006}. Therefore, the spectral evolution seen for
\swiftj\ is the expected at the observed X-ray luminosities, suggesting
that it is, at least, close to its intrinsic luminosity.
     
\subsection{Timing behaviour}

The broadband power density spectrum seen with both \xmm\, and \rxte\,
resembles, in both the shape as well as the strength of the
components, those seen from other black hole X-ray transients at low
luminosities. In particular, the power spectrum can be modeled with
either two Lorentzians or with a broken power-law model plus a
Lorentzian. In the latter case, the break frequency (0.038$\pm$0.002
Hz) and the frequency of the Lorentzian (0.33$\pm$0.06 Hz) obtained
using the \xmm\ data fall exactly on the relation found by
\citet[][]{1999ApJ...514..939W}, clearly identifying these components
with similar components seen in other black hole systems as well as in
neutron star X-ray binaries. However, contrary what is seen in those
neutron-star systems
\citep[e.g.,][]{2000A&A...358..617S,2007ApJ...660..595L}, we do not
see power at frequencies higher than those two components, which is
similar to what is seen in other black hole systems. This supports the identification that the accretor in this system is a
black hole and not a neutron star \citep[][]{Corral-Santana2013}.

As discussed previously, \citet[][]{Corral-Santana2013} suggested that
the source is viewed at very high inclination and an ADC
source. However, in our \xmm\ observation, and in the \rxte\ data as
well \citep[see also][]{Corral-Santana2013}, we do not see any signs
of the 2.8 hr orbit which would be expected for such high
inclinations. \citet[][]{Corral-Santana2013} proposed that the absence
of any orbital period signature in the X-ray data could be due to the
fact that the system has an extreme mass ratio and therefore the very
small size of the donor, making an X-ray eclipse very shallow and
difficult to detect in \swiftj. It is unclear if this can indeed
explain the lack of any orbital signature in the \xmm\, data as well.

The hypothesis for a high inclination was also indicated by the
optical dipping behavior of the source \citep[][]{Corral-Santana2013}.
No such dipping is visible in the X-ray data obtained with \xmm. We
also searched the \rxte\, data and found that during the first
observation a quasi-periodic oscillation is present in the X-ray
data. The QPO is at similar
frequencies as the frequency of the optical dips, however, we only see
the QPO in the first \rxte\, observation and not in any of the
following observations. In particular, no oscillations were seen in
the \rxte\, observations closest to the dates of the detection of the
optical dips. In addition, the frequency of the optical dips decreased
with decreasing X-ray luminosity of the source. Extrapolating the
frequency towards the start of the outburst would give a much higher
frequency for the dips or QPO than we observed for the X-ray QPO (see
Fig.~\ref{fig:lc_qpo}, bottom panel). Therefore, it is not clear
whether the mechanism behind the optical dips is similar to that
behind the X-ray QPO, or whether we observe two distinct phenomena
which happen to produce variation at similar time scales. In the
former case, the frequency either first is anti-correlated with X-ray
luminosity but later correlated, or we might see different harmonics
in the X-rays compared to in the optical. Neither possibility is
straightforwardly explained by an inner disc torus moving outwards
when the X-ray luminosity decreases \citep[as suggested by][to explain
the optical dips]{Corral-Santana2013}.

Alternatively, the X-ray QPO we observed in \swiftj\ could be part of
the recently suggested new class of black hole QPOs
\citep[][]{2012ApJ...754L..23A}: a low frequency QPO with frequencies
in the mHz range which is only seen at the start of an outburst. Such
a QPO has so far only been seen in one system \citep[namely
H1743--322; see][]{2012ApJ...754L..23A} but the properties of the QPO
in that system resembles the QPO seen in \swiftj. Both QPOs are at low
frequencies (11 mHz versus 6 mHz for H1743--322 and \swiftj\,
respectively) and only seen at the start of an outburst\footnote{We
note that in H1743--322, when it was present, the QPO was seen only at
the start of an outburst, but not every outburst of the source shows
the QPO even if ample data were available;
\citet[][]{2012ApJ...754L..23A}} (although we  note that the upper limits on the QPO strength after the first observation of \swiftj\ are not very constraining so this QPO might have been present during some later observations as well). We suggest that the QPO seen in
\swiftj\, is indeed similar to the QPO seen in H1743--322, making
\swiftj\ the second black hole system in which this QPO is seen.

\citet[][]{2012ApJ...754L..23A} compared this "mHz QPO" with other
QPOs seen in black hole and neutron star X-ray binaries and they
concluded that it best resembled the so-called "1 Hz QPO" seen in
dipping (and thus at relatively high inclination) neutron star systems
\citep[][]{1999ApJ...511L..41J,2000ApJ...531..453J,1999ApJ...516L..91H,2006ApJ...639L..31B,2012ApJ...760L..30H},
although at significantly lower (a factor of 10-100) frequencies. This
1 Hz QPO could be due to a structure in the inner disc which
quasi-periodically obscures the inner region
\citep[][]{1999ApJ...511L..41J}, although recently it was proposed
that these QPOs might be due to relativistic Lense-Thirring precession
of the inner accretion disc \cite[][]{2012ApJ...760L..30H}.
\citet[][]{2012ApJ...754L..23A} suggested that the difference in
frequency of the mHz QPO and the 1 Hz QPO might be explained if the
frequency of the QPO scales with the mass, or that it might depend on
the orbital period of the system (the orbital period of H1743--322
might be significantly longer than the orbital period of the dipping
neutron star systems which exhibited this QPO). However, if the QPO in
\swiftj\, is due to the same mechanism as the QPO in H1743--322 and if
they are related to the 1 Hz QPOs, then the orbital period is not
relevant because the orbital period of \swiftj\, is only 2.8 h,
leaving only the mass of the accretor as a potential determinant for
the exact observed QPO frequency.

We note that it cannot be conclusively stated that the X-ray QPO we
see in \swiftj\, is indeed related to the 1 Hz QPO in the dipping
neutron star systems, however, it is intriguing that a high
inclination was also inferred \citep[][]{Corral-Santana2013} for
\swiftj, similar to the dipping neutron star systems. This then would
support the hypothesis that the X-ray QPO in \swiftj\, might still be
related to the optical dips seen from the systems, although it is
currently not clear how exactly all observed phenomena can be
explained consistently.

\section{Final remarks}

The X-ray spectral and timing properties we have observed for
\swiftj\, in combination with the optical dips seen in outburst by
\citet[][]{Corral-Santana2013}, presents us with a complex range of
phenomena which needs to be explained. From the X-ray and optical
timing behavior (in combination with the broad width of the observed
H$_\alpha$ line) it was suggested that the source is at a relatively
high inclination. \citet[][]{Corral-Santana2013} suggested even such a
high inclination that we see the source edge on and the X-rays might
then only be due to scattered light (originating from the inner region
near the black hole) in an ADC. This might explain the optical dips,
the low observed X-ray luminosities (intrinsically the luminosity
would be much higher) and the broad H$_\alpha$ profile
\citep[][]{Corral-Santana2013}. However, the lack of orbital
modulation in the X-rays (neither eclipses nor the characteristic
X-ray dips were seen) and the lack of strong emission or absorption
lines in the X-ray spectrum obtained with \xmm\, cast doubt on this
edge-on hypothesis. In addition, a lower inclination might also allow
for an alternative (at least partly) explanation (then the one
suggested by \citealt[][]{Shahbaz2013}) why the orbital modulation in
the companion star flux could not be detected (lower inclination gives
rise to lower amplitude of this modulation) in quiescence and why in
quiescence the H$_\alpha$ line was not double peaked (lower
inclination gives rise to less separated lines). Furthermore, we see
clearly a soft thermal component at low energies in the \xmm\,
spectrum which probably originates from the inner edge of the disc,
also strongly indicating that we can observe directly the inner part
of the system. Therefore, we suggest that the system is still at
relatively high inclination so that the obscuring region which causes
the optical dips and the X-ray QPO still intersect periodically our
line of sight to the inner part of the system, but that the
inclination is not high enough to permanently obscure our view from
the inner part. This would mean that the luminosity we observe from
\swiftj\ is its intrinsic luminosity and therefore the source is
indeed a VFXT. Why the source has such a low intrinsic luminosity and
how exactly the optical dips and X-ray QPO are produce remains to be
determined.

\section*{Acknowledgments}

We acknowledge the \xmm\ team for make this observation possible. RW
and MAP are supported by an European Research Counsil starting grant
awarded to RW. ND is supported by NASA through Hubble Postdoctoral
Fellowship grant number HST-HF-51287.01-A from the Space Telescope
Science Institute (STScI).

\bibliographystyle{mn2e}

\input{references.tex}

\label{lastpage}
\end{document}

%% file: def_bibtex.tex
\def\apj{ApJ}
\def\mnras{MNRAS}
\def\aap{A\&A}
\def\apjl{ApJL}
\def\pasj{PASJ}

\def\astl{Astron. Lett. }

\def\ssr{Space Sci. Rev.}

\def \mnras {MNRAS}
\def \apj {ApJ}

\def \apjl {ApJL}
\def \aap {A\&A}

\def \pasj {PASJ}

%% file: definitions.tex
\def\lesssim{\mathrel{\hbox{\rlap{\hbox{\lower4pt\hbox{$\sim$}}}\hbox{$<$}}}}
\let\la=\lesssim
\def\gtrsim{\mathrel{\hbox{\rlap{\hbox{\lower4pt\hbox{$\sim$}}}\hbox{$>$}}}}
\let\ga=\gtrsim

\def\arcsec{\hbox{$^{\prime\prime}$}}

\def\sol{~\mathrm{M}_\odot}
\def\lx{L_\mathrm{X}}

\def\Fx{$F_\mathrm{X}$}

\def\Nh{$N_{\rm H}$}

\def\chired{$\chi^{2}_{\nu}$}
\def\chis{$\chi^{2}$}

\newcommand{\swift}{\textit{Swift}}
\newcommand{\xmm}{\textit{XMM-Newton}}
\newcommand{\rxte}{\textit{RXTE}}

\newcommand{\lum}{\mathrm{erg~s}^{-1}}
\newcommand{\flux}{\mathrm{erg~cm}^{-2}~\mathrm{s}^{-1}}
\newcommand{\cnts}{\mathrm{counts~s}^{-1}}
\newcommand{\nh}{\mathrm{cm}^{-2}}